\documentclass[twocolumn,epsfig,graphics,mathbbm,nopacs,hyperref]{revtex4-1}
\usepackage{amsmath,amsfonts,amssymb,amsthm,graphics,graphicx,epsfig,times,bbm}
\usepackage{amsmath}
\usepackage{amssymb}
\usepackage{graphics}
\usepackage{graphicx}

\newtheorem{observation}{Observation}
\newtheorem{lemma}{Lemma}

\usepackage{amsfonts}
\usepackage{psfrag}
\usepackage{dsfont}
\usepackage[dvips]{rotating}
\usepackage{epsfig}
\usepackage{pstricks,pst-grad}

\newcommand{\cc}{\mathbbm{C}}

\newcommand{\rr}{\mathbbm{R}}
\newcommand{\id}{\mathbbm{1}}

\begin{document}
\unitlength=1mm

\title{Limitations of quantum computing with Gaussian cluster states}

\date{\today}
\author{M.\ Ohliger${}^1$, K.\ Kieling${}^1$, and J.\ Eisert${}^{1,2}$}
\affiliation{${}^1$Institute of Physics and Astronomy, University of Potsdam, 14476 Potsdam, Germany}
\affiliation{${}^2$Institute for Advanced Study Berlin, 14193 Berlin, Germany}

\begin{abstract}
We discuss the potential and limitations of Gaussian cluster states for measurement-based 
quantum computing. Using a framework of Gaussian projected entangled pair states (GPEPS),
we show that no matter what Gaussian local measurements are performed on systems distributed on a general graph,
transport and processing of quantum information is not possible beyond a certain influence
region, except for exponentially suppressed corrections. We also demonstrate that even under
arbitrary non-Gaussian local measurements, slabs of Gaussian cluster states of a finite width
cannot carry logical quantum information, even if sophisticated encodings of qubits in 
continuous-variable (CV) systems are allowed for. This is proven by suitably contracting tensor networks representing 
infinite-dimensional quantum systems. 
The result can be seen as sharpening the requirements for quantum error correction and fault tolerance
for Gaussian cluster states, and points towards the necessity of non-Gaussian resource states for measurement-based quantum
computing. The results can equally be viewed as referring to Gaussian quantum repeater networks.
\end{abstract}
\maketitle

\section{Introduction}

Optical systems offer a highly promising route to quantum information processing and quantum computing. 
The seminal work of Ref.\ \cite{klm} showed that even with linear optical gate arrays alone 
and appropriate photon counting measurements, efficient linear optical computing is possible. 
The resource overhead of this proof-of-principle architecture for quantum computing was
reduced, indeed by orders of magnitude, by directly making use of the idea of measurement-based
quantum computing with cluster states \cite{clusterbriegel,tez,konradminimalresources}. Such an approach
is appealing for many reasons, the reduction of resource overheads only being one, but also for a
clearcut distinction between creation of entanglement as a resource and its consumption in computation.
This idea was further developed into the continuous-variable (CV) version thereof 
\cite{gaussiancluster,gaussiancluster2,gaussiancluster3,gaussiancluster4}, aiming at avoiding 
limitations related to efficiencies of creation and detection of single photons. In this context, Gaussian states play
a quite distinguished role as they can be created by passive optics, optical squeezers and coherent states, i.e., 
the states produced by a usual laser \cite{chracterizegaussian, transformpuregaussians,cvreview2,cvreview,cvreview3}: Indeed, 
Gaussian cluster states  form a promising resource for instances of quantum computing with light.
Such a CV-scheme allows for a deterministic preparation of the resource states while the schemes based on linear optics with single photons require preparation methods which are intrinsically probabilistic.

\begin{figure}
\begin{center}
\includegraphics[width=7.4cm]{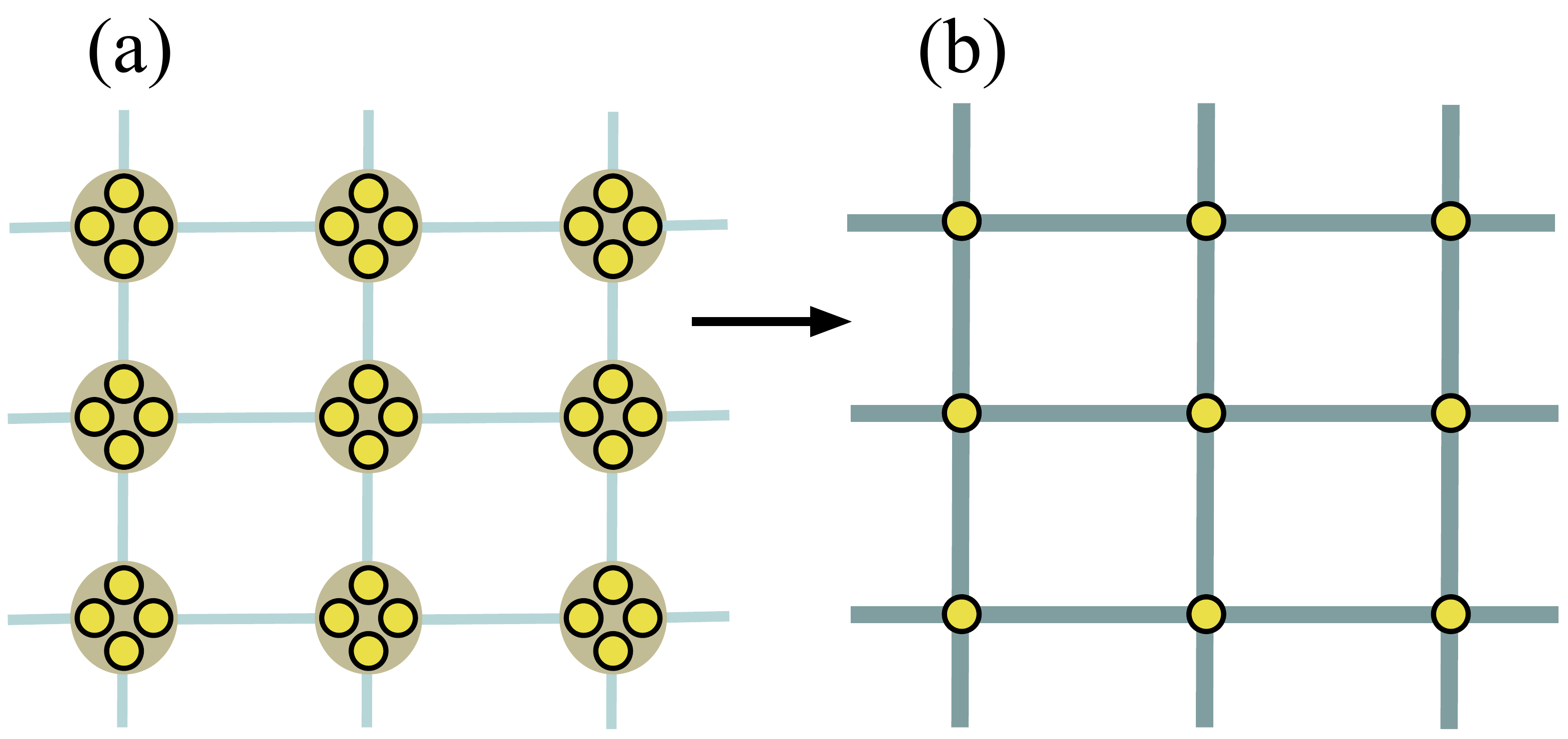}

\caption{\label{fig:gpeps}GPEPS on an arbitrary graph, here one representing a cubic lattice. (a) 
The connected dots represent two-mode squeezed states, the circles denote the vertices where the Gaussian projections 
are being performed. (b) The resulting GPEPS after local Gaussian projections have been performed
on the virtual systems. Any Gaussian cluster state can be prepared in this fashion.}
\end{center}
\end{figure}

In this work, we however highlight and flesh out some limitations of such an approach. We do so in
order to sharpen the exact requirements that any scheme for  CV quantum computing based on Gaussian 
cluster states eventually will have to fulfill, and what quantum error correction and fault tolerant approaches
eventually have to deliver. Specifically, we will show that
 Gaussian local measurements alone will not suffice to transport quantum information across the lattice,
 even on complicated lattices described by an arbitrary graph of finite dimension: 
 Any influence of local measurements is confined to a local region, except from exponentially suppressed corrections. 
 This can be viewed as an impossibility of Gaussian error correction in the measurement-based setting.
 What is more, even under non-Gaussian measurements, this obstacle cannot be overcome, 
 in order to transport or process quantum information along slabs of a finite width: Any influence of local measurements
 will again exponentially decay with the distance. This observation suggests that---although the initial state is 
 perfectly known and
 pure---finite squeezing has to be tackled with a full machinery of quantum error correction 
 and fault tolerance
 \cite{faulttolerantnielsen,faulttoleranttraussendorf,encoding}, 
 yet  to be developed for this type of system and presumably  giving rise to a massive overhead.  
 No local measurements or suitable sophisticated encodings of qubits in finite slabs---reminding, e.g., of
 encodings of the type of Ref.\ \cite{encoding}--- can uplift the initial state to an almost perfect 
 universal resource. In order to arrive at this conclusion, in some ways, we will explore ideas of 
 measurement-based computing beyond the one-way model \cite{clusterbriegel}
as introduced in Ref.\ \cite{prl} and further developed in Refs.\ \cite{pra,webs,Miyake,Chuang,BriegelSLOCC}.
We will highlight the technical results as ``observations'', and discuss implications of these results in the 
main text. While these findings do not constitute a ``no-go'' argument for Gaussian cluster states,
they do seem to require a very challenging prescription for quantum error correction and
further highlight  the need for identifying alternative schemes for CV quantum computing, specifically 
ones based on {\it non-Gaussian CV 
states}. {\it Small scale implementations} of {\textit Gaussian cluster state computing} are, as we will see, 
are also {\it not} affected by these limitations.

To sketch the structure of this article: In Section \ref{sec:GPEPS} we will first discuss the 
concept of Gaussian projected entangled pair states (GPEPS), forming a family of states including the physical CV Gaussian
cluster state.  In Section \ref{sec:GOPS} we will then discuss the impact of 
Gaussian measurements on GPEPS and show that under this restriction the localizable entanglement in every GPEPS 
decays exponentially with the distance between any two points on arbitrary lattice. This has also implications
on Gaussian quantum-repeaters, which we investigate in detail. After this, we leave the strictly Gaussian stage in Section \ref{sec:nogauss} 
and investigate present our main result when showing that under more general measurements on GPEPS, 
quantum information processing in finite slabs is still not possible. We discuss requirements for error correction, before presenting concluding 
remarks. 

\section{Preliminaries}

\subsection{Gaussian states}

Before we turn to measurement-based quantum computing (MBQC) on CV-states, 
we  briefly review some basic elements of the theory of Gaussian states and operations which are needed in this article  
\cite{chracterizegaussian, transformpuregaussians,cvreview,cvreview2}. Readers familiar with these concepts can safely skip this section.
Although the statements made in this work apply on all physical 
systems described by {\it quadratures} or
{\it canonical coordinates}, including, e.g., micromechanical oscillators, 
we have a {\it quantum optical system} in mind and often use language from this field as well. Any system of $N$ bosonic degrees of freedom, e.g., $N$ light modes, can be described by canonical coordinates $x_n=(a_n+a_n^\dag)/2^{1/2}$ and 
$p_n=-i(a_n-a_n^\dag)/2^{1/2}$, $n=1,\dots, N$, where $a_n$ ($a_n^\dag$) annihilates (creates) a photon in the respective mode. 
When we collect these $2N$ canonical coordinates in a vector $O=(x_1,p_1,\ldots,x_N,p_N)$, we can write the commutation 
relations as $[O_j,O_k]=i\sigma_{j,k}$, where the {\it symplectic matrix} $\sigma$ is given by
\begin{equation}
\sigma=\bigoplus_{j=1}^N\left[\begin{array}{cc}0 & 1 \\ -1 & 0\end{array}\right].
\end{equation}
{\it Gaussian states} are fully characterized by their first and second moments alone.
The {\it first moments} form a vector $d$ with entries
$d_j={\rm tr}(O_j\rho)$ while the second moments, which capture the fluctuations, can be 
collected in a $2N\times 2N$-matrix $\gamma$, the so-called {\it covariance matrix}, with entries
\begin{equation}
	\gamma_{j,k}=2{\rm Re}\,{\rm tr}\left[\rho\left(O_j-d_j\right)\left(O_k-d_k\right)\right].
\end{equation}
Hence, Gaussian states are complete characterized by $d$ and $\gamma$. {\it Gaussian unitaries}, i.e., 
unitary transformations acting 
in Hilbert space preserving 
the Gaussian character of the state correspond to symplectic transformations on the CM. They in turn correspond so maps
$\gamma\mapsto S\gamma S^T$ with $S\sigma S^T=\sigma$. The set of such {\it symplectic transformations}
forms the group $Sp(2N,\rr)$.
A set of particularly important example Gaussian states are the {\it coherent states}, for which the state vectors read in the photon number basis
\begin{equation}
|\alpha\rangle=e^{-|\alpha|^2/2}\sum_{n=0}^\infty\frac{\alpha^n}{\sqrt{n!}}|n\rangle
\end{equation}
and are described by $d=({\rm Re}\,\alpha,{\rm Im}\,\alpha)$ and $\gamma={\rm diag}(1,1)$. Single mode {\it squeezed states} 
are characterized by lower fluctuations in one phase-space coordinate. The CM can in a suitable basis then be written as $\gamma={\rm diag}(x,1/x)$ with $x\neq0$.  

\subsection{MBQC on Gaussian cluster states}

The first proposal for MBQC on CV states has been based on so-called {\it Gaussian cluster states} and works in almost complete analogy to the qubit case \cite{gaussiancluster,gaussiancluster2,gaussiancluster3,gaussiancluster4}. 
As such, the formulation is based on ``infinitely squeezed'' and hence unphysical states using infinite
energy in preparation: It can be created by initializing every mode in the $p=0$ ``eigenstate'' of $p$ (formally an improper
eigenstate of momentum, a concept that can be made rigorous, e.g., in an algebraic formulation \cite{cvw}).
This is the CV-analogue to the state vector
$|+\rangle=(|0\rangle+|1\rangle)/2^{1/2}$ in the qubit case. Then the operation $e^{ix\otimes x}$, the analogue to the 
$C_Z$ gate, is applied between all adjacent modes. This state allows universal MBQC to be performed 
with Gaussian and one non-Gaussian measurement. The state as such is not physical and not contained in Hilbert
space. The argument, however, is that it should be expected that 
a finitely squeezed version inherits essentially the same properties. 
Replacing them by finitely squeezed  
ones we obtain a state which we will call a {\it physical Gaussian cluster state}. 

\section{Gaussian PEPS}
\label{sec:GPEPS}
{\it Projected entangled pair states (PEPS)} or {\it tensor product states}
have been used for qubits to generalize {\it matrix product states (MPS)} 
or {\it finitely correlated states} \cite{MPStheoryhard,MPStheoryeasy}
from one-dimensional chains to arbitrary graphs 
\cite{PEPS,PEPS2,MBPEPS}. One suitable of defining them is via a 
valence-bond construction: One can create a state by placing entangled pairs---constituting ``virtual systems''---on 
every bond of the lattice and then applying a 
suitable projection to a single mode at every lattice site.
 These projections, often taken to be equal, together with the specification of the initial
 entangled states,  then serve as a description of the resulting state. {\it MPS for Gaussian states} (GMPS) have been 
 studied to obtain correlation functions and entanglement scaling
 in one-dimensional chains \cite{gaussianmps}. 
 
 In this work we focus on {\it Gaussian PEPS (GPEPS)} which can be obtained 
 from non-perfectly entangled pairs. The bonds we consider are {\it two-mode squeezed states 
 (TMSS)}, the state vectors of which  
 have the photon number representation
\begin{equation}\label{TMSS}
	|\psi_\lambda\rangle=(1-\lambda^2)^{1/2}\sum_{n=0}^\infty\lambda^n|n,n\rangle,
\end{equation}
where $\lambda\in (0,1)$ is the {\it squeezing parameter}. 
We will denote the corresponding density matrix by $\rho_\lambda$. For $\lambda\to 1$ the state becomes ``maximally entangled'',
but this limit is not physical because it is not normalizable and has infinite energy as already mentioned above. We will, 
therefore, carefully analyze the effects stemming from the fact that $\lambda<1$. The covariance matrix of this state reads
\begin{equation}
\label{eq:TMSS}
\gamma_\lambda=\left[\begin{array}{cccc}{\rm cosh}(2r)& 0 & {\rm sinh}(2r) & 0 \\ 0 & {\rm cosh}(2r) &0 &-{\rm sinh}(2r) \\ {\rm sinh}(2r) & 0 & {\rm cosh}(2r) & 0 \\ 0 & -{\rm sinh}(2r) & 0 & {\rm cosh}(2r)\end{array}\right]
\end{equation}
where ${\rm tanh}(r/2)=\lambda$. This number $r$ will also be referred to as the {\it squeezing parameter} in 
case there is no risk of mistaking one for the other. It is also known that any pure bi-partite multi-mode Gaussian state
can be brought into the tensor product of TMSS \cite{transformpuregaussians,alonso} by means of local unitary
Gaussian operations, each having a CM in the above form. 
Then the largest $r$ in the vector of resulting TMSS will be referred to as its squeezing parameter. 

We will also discuss GPEPS on general graphs $G=(V,E)$,
as shown in Fig.\ \ref{fig:gpeps}. Vertices $G$ here correspond to physical systems, edges $E$
to connections of neighborhood. On any such graph, $d(.,.)$ is the natural graph-theoretical distance
between two vertices. As we will often consider the system of bonds before the projection operation is performed, we employ the 
following notation: When we speak of operations on {\it virtual systems} when thinking of
collective operations on modes before the projection is applied, and often emphasize when 
we refer to a single physical system with Hilbert space ${\cal H}={\cal L}^2(\rr)$. Note that we also allow 
for more than one edge between two vertices of a graph. 

When a particular vertex has $N$ adjacent bonds, the projection map is a Gaussian operation of the form
\begin{equation}
\label{eq:prepmap}
V:{\cal H}^{\otimes N}\rightarrow{\cal H}.
\end{equation}
This operation can always be made trace-preserving \cite{cvreview,nogaussiandistill,chracterizegaussian,fiurasek},
in quite sharp contrast to the situation in the finite-dimensional setting. This operation will also be referred
to as Gaussian {\it PEPS projection}. This operation can always be realized by mixing single-mode squeezed state on a suitably tuned beam splitter which means that \textit{inline} squeezers are not necessary \cite{passgeneration}. 
Note that any such state could also be used as a variational
state to describe ground states of many-body systems 
and by construction satisfies an {\it entanglement area law} \cite{area}.

\section{Gaussian operations on GPEPS}
\label{sec:GOPS}

In this section, we will discuss Gaussian operations on a GPEPS 
and derive some statements on entanglement swapping, the localizable entanglement, and the usefulness as a 
resource for MBQC. Since all measurements are assumed to be Gaussian as well, this is as such not yet
a full statement on universality, but already shows that the natural operations for transport of logical information
in such a Gaussian cluster state does not work with such local measurements.

\subsection{Localizable entanglement}
\label{sus:LE}
The {\it localizable entanglement (LE)} 
between two sites $A$ and $B$ of the graph $G=(V,E)$
is defined by the maximal entanglement obtainable on average when performing projective measurements at all sites but $A$ and $B$ \cite{localizableentanglement}. When we require both the initial state and the measurements to be Gaussian \cite{gaussianle,gaussianlemixed}, 
the situation simplifies as the entanglement properties do not depend on the measurement outcomes \cite{cvreview,nogaussiandistill,chracterizegaussian,fiurasek}. 
Thus, we do not need to average but only find the best measurement strategy. To be specific, we will 
measure the entanglement in terms of the {\it logarithmic 
negativity} which can be defined as \cite{negativity0,negativity,negativity2}
\begin{equation}
	E(\rho)={\rm log}\|\rho^{T_A}\|_1,
\end{equation}
where $T_A$ denotes the {\it partial transpose} with respect to subsystem $A$ and $\|.\|_1$ the trace-norm and we use the natural logarithm.
For a TMSS, $E$ coincides with the squeezing parameter by $E(\rho_\lambda)=r$. It is important
to note, however, that this choice has only been made for notational convenience: In our
statements on asymptotic degradation of entanglement, any other measure of entanglement would also do, specifically
the entropy of entanglement for pure Gaussian states, and for mixed states the 
{\it distillable entanglement} or the {\it entanglement cost}.

We will mostly focus on two variants of the concept of localizable entanglement:
Whenever we only allow for Gaussian local measurements,
we will refer to this quantity as {\it Gaussian localizable entanglement}, abbreviated as $E_{\rm G}$.
Then, we will consider the situation when we ask for fixed subspaces $S_A$ and $S_B$ in the Hilbert
spaces associated with sites $A$ and $B$ to get entangled by means of local measurements.
We then refer to as {\it subspace localizable entanglement} $E_{\rm S}$.  Both concept directly
relate to transport in measurement-based quantum computing.

\subsection{Entanglement swapping}
\label{sus:onedgpepsg}

The task of localizing entanglement in a PEPS is closely related to one of {\it entanglement 
swapping} \cite{loockpdhthesis}. In this situation we have three parties, $A$, $B$, and $C$ 
where both $A$ and $B$ and $B$ and $C$ share an entangled pair each. 
Then $B$, consisting of  $B_1$ and $B_2$, 
is allowed to perform an arbitrary Gaussian 
operation on his parts of the two pairs followed by a measurement. The task is to choose the operation in 
such a way that the resulting entanglement between $A$ and $B$ is maximum. 

\begin{lemma}[Optimality of Gaussian Bell measurement for entanglement swapping of two-mode squeezed states] 
For two pairs of entangled
TMSS shared between $A$ and $B_1$ and $B_1$ and $C$, the supremum of maximum achievable 
negativity between $A$ and $C$ by a local Gaussian measurement in $B_1,B_2$ is approximated
by the measurement that best approximates a Gaussian Bell measurement.
\end{lemma}

\begin{figure}
\begin{center}
\includegraphics[width=4.4cm]{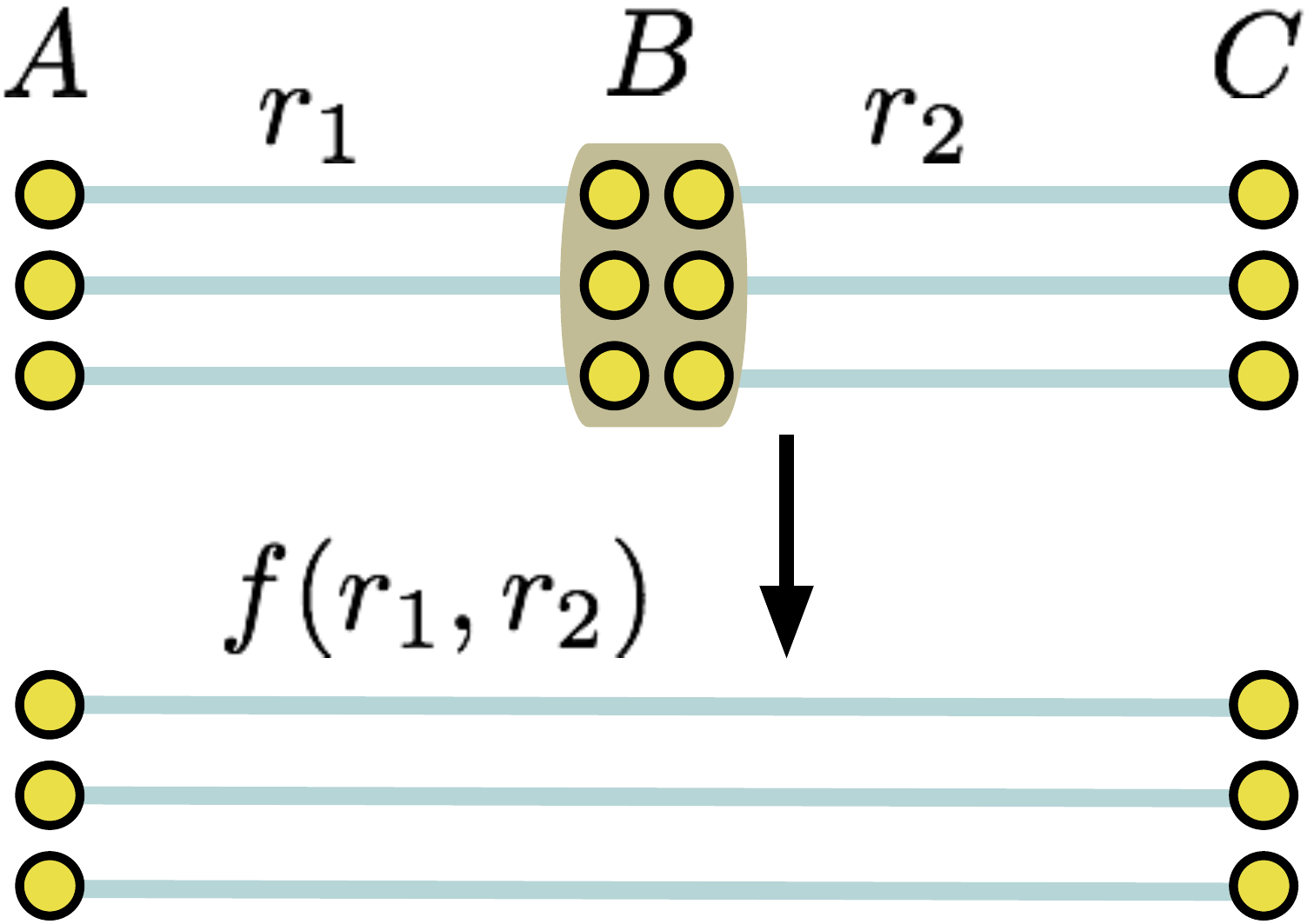}
\caption{\label{fig:nocolective}Situation referred to in Lemma \ref{lem:nocolective}. The strongest 
bonds before the projection are $r_1$ and $r_2$. The most significantly entangled bond has the strength  $f(r_1,r_2)$.}
\end{center}
\end{figure}
We consider the situation of having a TMSS 
(\ref{eq:TMSS}) 
\begin{equation}
	|\psi\rangle_{A,B_1}=|\psi_{\lambda_1}\rangle_{A,B_1},\,\,
	|\psi\rangle_{B_2,C}=|\psi_{\lambda_2}\rangle_{B_2,C}
\end{equation}	
with some $\lambda_1,\lambda_2>0$
and restricting the operation on $B$ to be Gaussian. Furthermore, we allow for operations which do not succeed with unit probability. We have to allow for general local Gaussian operations, and also for arbitrary local additional Gaussian resources, with CM $\gamma_B$ on modes $B_3$,
on an arbitrary number of modes.
The initial covariance matrix of the system hence reads 
\begin{equation}	
	\gamma=\gamma_{\lambda_1}\oplus\gamma_{\lambda_2}\oplus \gamma_{B_3}. 
\end{equation}	
Without loss of generality, one can assume 
that one performs a single projection onto a pure Gaussian state on all modes referring to $B$. Ordering 
modes to $A$, $C$, $B_1$, $B_2$, $B_3$, one can write the CM in block form as
\begin{equation}
	\gamma=\left[\begin{array}{ccc}U & V & 0 \\ V^T & W & 0\\
	0 & 0 & \gamma_{B_3}\end{array}\right] ,
\end{equation}
$U$ referring to $A$, $C$ and $V$ to $B_1$, $B_2$. 
When we now project the modes $B_1$, $B_2$, and $B_3$ onto a pure Gaussian state with 
CM $\Gamma$, the CM of the resulting state of $A$ and $C$, postselected on that outcome,  
is given by the 
Schur-complement \cite{nogaussiandistill,chracterizegaussian,fiurasek},
\begin{equation}
\label{eq:schurcomplement}
	\gamma_{A,C}=\left[
	\begin{array}{cc}
	U & 0\\
	0 & 0\\
	\end{array}\right]
	-
	\left[\begin{array}{cc}
	V & 0\\
	\end{array}\right]
	\left(
	\left[\begin{array}{cc}
	W & 0\\
	0 & \gamma_{B_3}
	\end{array}
	\right]
	+\Gamma\right)^{-1}
	\left[\begin{array}{c}
	V^T\\ 0\\
	\end{array}\right].
\end{equation}
Any symplectic operation $S$ applied to $B$ before the measurement
can of course also be just absorbed into 
the choice of the CM $\Gamma$.
Writing 
\begin{equation}
\left[\begin{array}{cc}
	W & 0\\
	0 & \gamma_{B_3}
	\end{array}
	\right] + \Gamma = \left[\begin{array}{cc}
	X & Y\\
	Y^T &  Z
	\end{array}
	\right],
\end{equation}
one finds that the left upper principal submatrix of the 
inverse can be written as
\begin{equation}
	\left.\left[\begin{array}{cc}
	X & Y\\
	Y^T &  Z
	\end{array}
	\right]^{-1}\right|_{B_1,B_2} = (X-Y Z^{-1}Y^T)^{-1} ,
\end{equation}
again in terms of a Schur complement expression. Since $\gamma_{B_3}+i\sigma\geq 0$ and the same
holds
for the subblock on $B_3$ of $\Gamma$, these matrices are clearly positive. Using operator monotonicity of the 
inverse function, one finds that 
\begin{equation}
	(X-Y Z^{-1}Y^T)^{-1} \geq 0 
\end{equation}
holds, since $Y Z^{-1}Y^T \geq 0$. Therefore,
\begin{equation}
	\gamma_{A,C} = \gamma_{A,C}' +P
\end{equation}
with a matrix $P\geq 0$. Here $\gamma_{A,C}'$ is the CM following the same protocol, but where
$\Gamma$ is replaced by the identical CM, but with $Y=0$. To arrive at such a CM 
is always possible and still gives rise to 
a valid CM by virtue of the pinching inequality. This is yet merely the covariance matrix of the
a Gaussian state, subjected to additional classically correlated Gaussian noise. 
In other words, it is always optimal to treat $B_3$
as an innocent bystander and not to perform an entangling measurement between $B_1$ and $B_2$
on the one hand and $B_3$ on the other hand, quite consistent with what one could have intuitively
assumed. We can hence focus on the situation when $B_3$ is absent and we merely
project onto a pure Gaussian state in $B_1$ and $B_2$.

It is then easy to see that 
there is no optimal choice, but the supremum can be better and
better approximated by considering more and more squeezed TMSS (or ``infinitely squeezed states'' in the first place),
i.e.,  
on $|\psi_\lambda\rangle$ in the limit of $\lambda\rightarrow 1$,
which is the CV-analogue to the Bell state for qudits. This measurement can be realized by mixing $B_1$ and $B_2$ on a beam splitter with reflectivity $R=1/2$ and performing homodyne measurement on both modes afterwards (i.e., a projection on a infinitely squeezed single-mode state
being improper eigenstates of the position operator). From Eqs.\ (\ref{eq:TMSS}) and (\ref{eq:schurcomplement}) with $\Gamma=\gamma_\lambda$ and performing the limit $\lambda\to 1$, we can calculate the CM of the resulting state. It has the form of (\ref{eq:TMSS}) with
\begin{equation}
	\label{eq:fr}
	r=f(r_1,r_2)=\frac{1}{2}{\rm arcosh}\frac{1+{\rm cosh}2r_1{\rm cosh}2r_2}{{\rm cosh}2r_1+{\rm cosh}2r_2}.
\end{equation}
We note that $f$ is symmetric in its arguments and fulfills $f(r_1,r_2)<{\rm min}\{r_1,r_2\}$ and $\lim_{r_1\to\infty}f(r_1,r_2)=r_2$. 
This means, arbitrarily 
faithful entanglement swapping is possible exactly in the limit of
infinite entanglement. Otherwise the entanglement necessarily deteriorates \cite{loockpdhthesis}.

To show that this measurement is indeed optimal, we set 
\begin{equation}
	\Gamma=S\gamma_\lambda S^T
\end{equation}
 where $S \in Sp(4,\rr)$. 
Calculating the resulting degree of entanglement, a direct and straightforward inspection reveals that
$E(\rho_{A,C})$ can only decrease whenever 
we choose $S\neq\id$. 

\subsection{One-dimensional chain}
\label{sus:oned}

We now turn to a one-dimensional GPEPS, not allowing multiple bonds in the valence-bond construction, 
and are in the position to show the following observation: 

\begin{observation}[Exponential decay of Gaussian localizable entanglement in a 1D chain]
\label{the:oned}
Let $G$ be a one-dimensional GPEPS and $A$ and $B$ two sites. Then
\begin{equation}
	E_{\rm G}(A,B)\le c_1 e^{-d(A,B)/\xi_1 }
\end{equation}
where 
$c_1,\xi_1>0$ are constants. The best performance is reachable by passive optics and homodyning only. 
\end{observation}

In order to prove this, we interpret the preparation projection (\ref{eq:prepmap}) and the following measurements of the 
localizable entanglement protocol as a sequence of instances of entanglement swapping. Clearly, to allow for general
Gaussian projections is more general than (i) using the specific Gaussian projection of the PEPS, followed by
a (ii) suitable Gaussian projection onto a single mode; hence every bound shown for this setting will also
give rise to a bound to the actual 1D Gaussian chain. If $d(A,B)$ is again
the graph-theoretical distance between $A$ and $B$ we have to swap $k=d(A,B)-1$ times. 
Defining $g(r)=f(r,r_I)$ where $r_I$ is the initial strength of all bonds and iterating the argument we obtain 
\begin{equation}
r_{A,B}=(g^{\circ k})(r_I)=F(k)\quad.
\end{equation}
As the negativity is up to a simple rescaling 
equal to this two-mode-squeezing parameter, the only task left is to show that $F(k)$ decays exponentially. 
In order to do this, we need ${\rm arcosh}(x)=\log(x+(x^2-1)^{1/2})$ and the following relations 
which hold for $x\ge 0$: ${\rm cosh}(x)\ge e^x/2$, ${\rm cosh}(x)\le e^x$. With the help of them, we can conclude that
\begin{equation}
	F(k+1)/F(k)<Q<1
\end{equation}	
for a $Q$ only depending on $r_I$. Thus, $F(k)$ decays exponentially which proves Observation $\ref{the:oned}$. 
Note that in order to maximize the entanglement between $A$ and $B$, we have chosen the 
supremum of the
maps better and better approximating the projection onto an infinitely entangled TMSS. 
Thus, for a specific GPEPS which is characterized by a fixed map $V$, the $E_G$ in generally lower. 

This result has a remarkable consequence for Gaussian {\it quantum repeaters lines}: 
It is not possible to build a one-dimensional quantum repeater relying on Gaussian 
states, if only local measurements and no distillation steps are being used. 
We will show in Section \ref{sec:nogauss} that even non-Gaussian measurements 
cannot improve the performance. If one 
sticks to the Gaussian setting, also relying on complex networks does not remedy 
the exponential decay, as we will see. Of course, non-Gaussian distillation schemes can be 
used in order to realize CV quantum repeater networks.

\subsection{General graphs in arbitrary dimension}

One should suspect that the exponential decay of $E_{\rm G}$ is a special feature of the 
one-dimensional situation and that higher dimensional graphs would eventually 
allow to localize a constant amount of entanglement. In this section we will show 
that this is not the case. We first need a Lemma which follows directly 
from our discussion of entanglement swapping.

\begin{lemma}[Collective operations on pure Gaussian states]
\label{lem:nocolective} Let $\rho_{A,B_1}$ be a pure Gaussian
state on ${\cal H}^{\otimes 2 n}$ of $n$ modes
and $\rho_{B_2,C}$ a pure Gaussian ${\cal H}^{\otimes 2 m}$
state, where one part of each is held by $A$, $B$,  and $C$, respectively. Let the 
 maximum two-mode squeezing parameter be $r_1$ between 
 $A$ and $B$ and $r_2$ between $B$ and
 $C$.  Then the maximum two-mode-squeezing parameter 
 achievable with a
 Gaussian projection in $B$ between $A$ and $C$ is $f(r_1,r_2)$.
\end{lemma}

\begin{figure}
\begin{center}
\includegraphics[width=4.8cm]{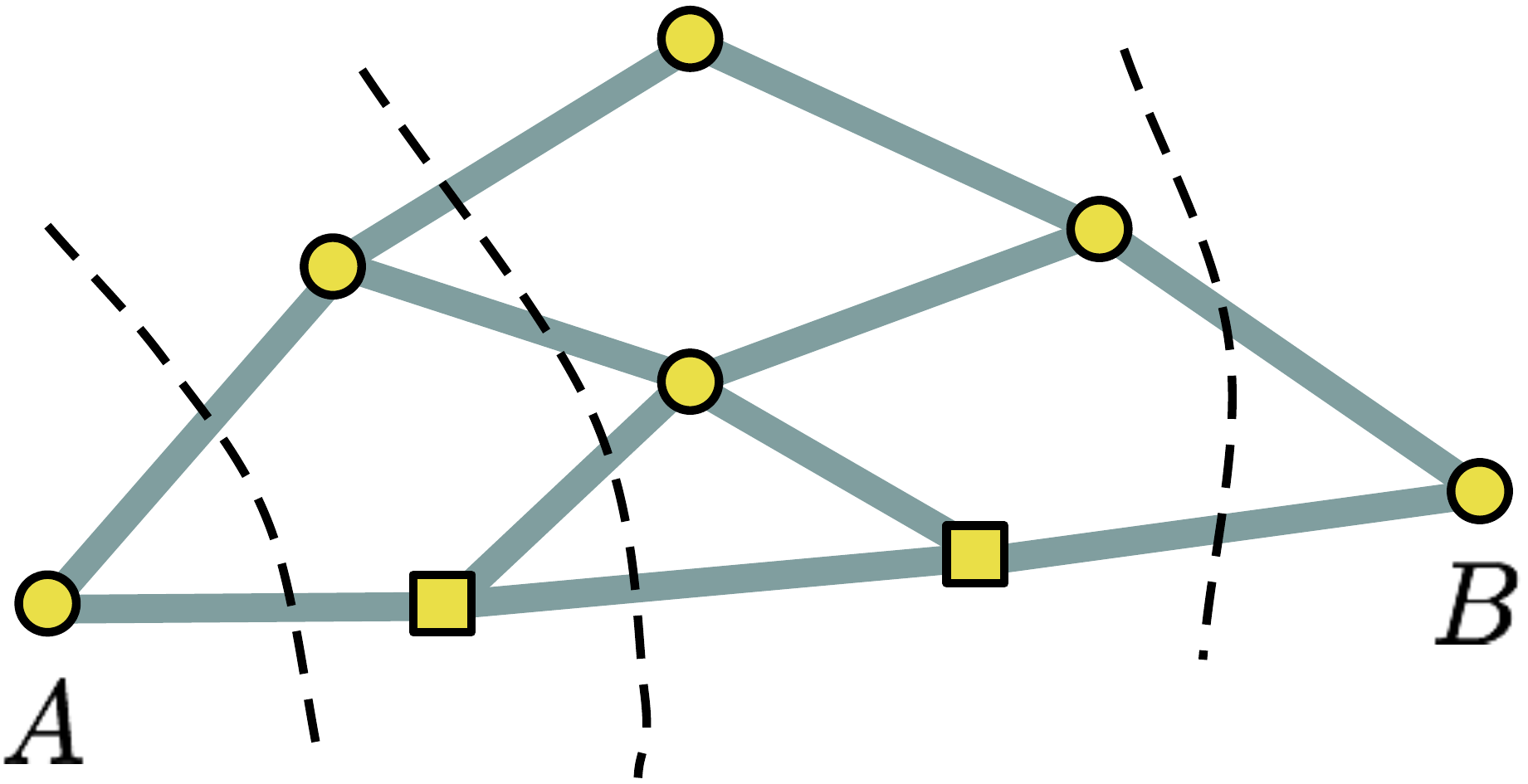}
\caption{\label{fig:slicing}Partitioning of the graph according to the shortest path as described in the main text. 
The sites drawn as squares are the ones which lie on the shortest path connecting $A$ and $B$.}
\end{center}
\end{figure}

To prove this, we again use the fact that any two-party multi-mode pure 
Gaussian state can be transformed by local unitary Gaussian operations on both parties into 
a product of TMSS \cite{transformpuregaussians,alonso}. 
This is nothing but the Gaussian version of the {\it Schmidt decomposition}. 
It does hence not restrict generality to start from that situation. 
As noted above, the best strategy for entanglement-swapping between two pairs is a Gaussian Bell-measurement, 
where the squeezing parameter changes according to $f$. 

We will now allow for global Gaussian operations on all
subsystems belonging to $B$. This situation we will relax to the following, where we
allow for even more general operations: Namely
a local Gaussian operation onto all modes of $B$, as well as onto all modes of $A$ and $C$ that are not
the two modes that share the largest $r$. Clearly, this is a more general map than is actually considered
in the physical situation. This, however, is exactly the situation considered above, of an entanglement
swapping scheme with an unentangled bystander. Hence, we again find that to project each pair
onto a two-mode pure Gaussian state is optimal. For that, the sequence of projections better and
better approximating infinitely squeezed TMSS gives rise to the supremum. Hence, 
we have shown the above result. 
Now we can prove a central result of this work.

\begin{observation}[Exponential decay of Gaussian localizable entanglement of GPEPS on general graphs]
\label{the:generalgraph}
Consider a GPEPS on a general graph with finite dimension
and let $A$ and $B$ be two vertices of this graph. Then there exist constants 
$c_2,\xi_2>0$ such that
\begin{equation}
	E_{\rm G}(A,B)\le c_2 e^{-d(A,B)/\xi_2}.
\end{equation}
\end{observation}

We take the shortest path between $A$ and $B$---achieving the graph-theforetical distance $d(A,B)$---and 
denote its vertices by $A,v_1,\ldots,v_{d(A,B)-1},B$. We partition the graph in such a way that the boundaries do not intersect or touch each other and every vertex on the shortest path from $A$ and $B$  is contained in one region which is called $R_v$ (see Fig~\ref{fig:slicing}). 
Again we consider the situation of having TMSS distributed on the graph between vertices
sharing an edge: A general local Gaussian
measurement on a Gaussian PEPS---so the Gaussian PEPS projection now on several modes, 
followed by a specific single-mode Gaussian measurement can only be 
less general than a general collective Gaussian measurement, so again we will arrive at a bound
to the localizable entanglement in the Gaussian PEPS.

Now, we face exactly the situation to which Lemma \ref{lem:nocolective} applies. In fact, we will
in each step in each of the parts $A$, $B$, and $C$ have a collection of TMSS, shared across the
cut of the three regions. 
If $r_{Av_1}$ is the strongest bond, in terms of the two-mode squeezing parameter, 
between $R_A$ and $R_{v_1}$ and $r_{v_1v_2}$ the strongest one 
between $R_{v_1}$ and $R_{v_2}$, 
then the strongest bond between $R_{A}$ and $R_{v_2}$ is given according to Lemma \ref{lem:nocolective} 
by $f(r_{Av_1},r_{Av_2})$. Now, we can proceed exactly as in the proof of Theorem \ref{the:oned}---and again
any uncorrelated bystanders will not help to improve the degree of entanglement---and 
thus show Theorem \ref{the:generalgraph}. This has again a consequence for 
quantum repeaters: Even when an arbitrary number of parties can 
share arbitrary many Gaussian entangled bonds, 
it is not possible to teleport quantum information over an arbitrary distance as shown below.

\begin{figure}
\begin{center}
\includegraphics[width=4.7cm]{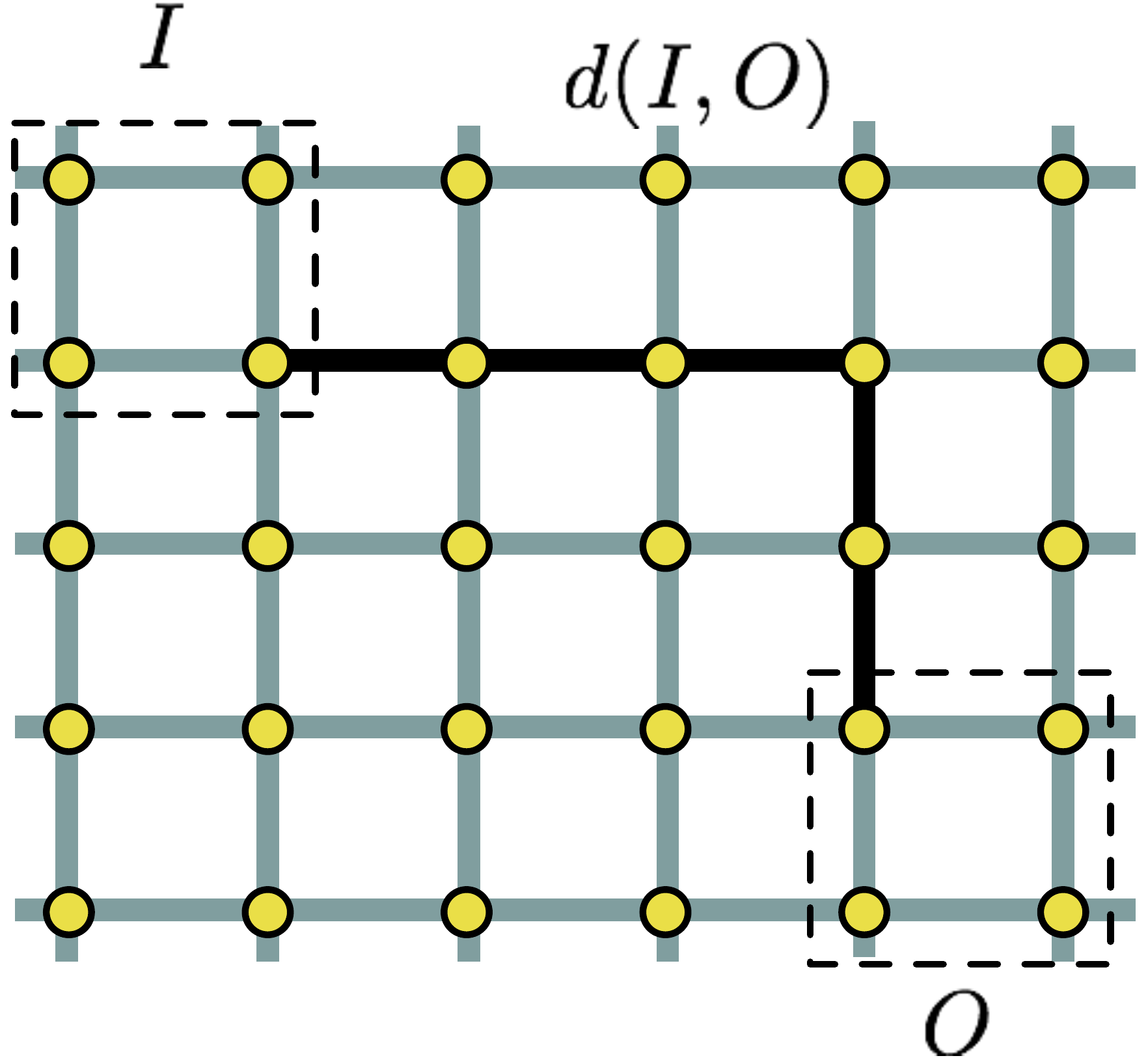}
\caption{\label{fig:decay}Exponential decay of any influence of any measurements of measurements in region $I$ to
statistics of measurement outcomes in $O$ in the graph theoretical distance $d(I,O)$ between the regions.}
\end{center}
\end{figure}

In fact, using this statement, one can show that any impact of measurements in terms of a measurable
signal in confined to a finite region on the graph, now a $I$ being a subset of the graph, 
expect from exponentially suppressed corrections. This region could be a poly-sized region in which
the input to the computation is encoded. The read-out of the quantum computation is then estimated from
measurements on some region $O$, giving rise to a bit that is the result of the original
decision problem that is to be solved by the quantum computation. 
From the decay of localizable entanglement, it is not difficult to show that the probability
distribution of this bit is unchanged by measurements in $I$, except from corrections that
are exponentially decaying with $d(I,O)$, see Fig.\ \ref{fig:decay}.

Note that concerning small scale, ``proof-of-principle'' applications, the presented arguments do {\it not} impose a fundamental 
restriction as they only apply to the situation where entanglement distribution over an arbitrary number of 
modes (or repeater stations) is required. For any \textit{finite} distance 
$d(A,B)$ and required entanglement $E(A,B)$ there exists a \textit{finite} 
minimal squeezing $\lambda_{\rm min}$ which allows to perform the task. Only asymptotically, one will necessarily
encounter this situation. The result can equally be viewed as an impossibility of Gaussian quantum 
error correction in a measurement-based setting, complementing the results of Ref.\ \cite{Cerf}.

\subsection{Remarks on Gaussian repeater networks}

These results of course also applies to general {\it quantum repeater networks}, where the aim is to end
up with a highly entangled pair between any two points in the repeater network 
(see, e.g., Ref.\ \cite{spinrep} for a qubit version thereof). That is, in Gaussian repeater networks, one will also need
non-Gaussian operations to make the network work, quite consistent with the
findings of Refs.\ \cite{nogaussiandistill,chracterizegaussian,fiurasek}. 

\subsection{Measurement based quantum computing}
\label{sus:MBQCgauss}
The impossibility of encountering a localizable entanglement that is not exponentially
decaying does directly lead to a statement on the impossibility of using a GPEPS as a quantum wire. 
Such a wire should be able to perform the following task \cite{prl}: 
Assume that a single mode holds an unknown qubit in an arbitrary encoding, i.e.,
\begin{equation}
\label{eq:encoding}
	|\phi_{\rm in}\rangle=\alpha|0_L\rangle+\beta|1_L\rangle
\end{equation}	
This system is then coupled to a defined site $A$, the first site of the wire, of a GPEPS by a fixed in-coupling unitary operation which can in general be non-Gaussian. To complete the in-coupling operation, the input mode is measured in an arbitrary basis, where we also allow for probabilistic protocols, i.e. the operation does not have to succeed for all measurement outcomes. 
Then one performs local Gaussian measurements on each of the modes. Then, at the end, 
one expects the mode at a single site $B$ to be in the state vector
$|\phi_{\rm out}\rangle=U|\phi_{\rm in}\rangle$ 
(or at least arbitrarily close in trace norm) for any chosen $U\in SU(2)$. 
Note that the length of the computation, and, therefore, the position of the output mode $B$, may 
vary and that the computational subspace can be left during the measurement. We want to 
stress that it is also possible to consider quantum wires which process qudits or even CV 
quantum information, where even on the logical level information is encoded continuously. 
However, the capability to process a qubit is clearly the weakest requirement. Thus, we 
will only address this situation because the corresponding statements for other quantum 
wires immediately follow. With this clarification we can state the following lemma.

\begin{observation}[Impossibility of using Gaussian operations on arbitrary GPEPS on general graphs for quantum wires]
\label{lem:nowire}
No GPEPS on any graph together with Gaussian measurements can serve as a perfect
quantum wire for even a single qubit.
\end{observation}

This is obvious from the previous considerations, as the measurements for the localizable entanglement
and the incoupling operation commute, and clearly, the procedure is especially not possible for 
$U=\id$. The same argument of course also holds true on general graphs, that no wire can be 
constructed from local Gaussian measurements in this sense, again for an exponential decay
of the localizable entanglement. As mentioned before, this statement can also be refined to having
up to exponential corrections finite influence regions altogether.

%%%%%%%%%%%%%%%%%%%%%%%%%%%%%%%%%%%

\section{Non-Gaussian operations}
\label{sec:nogauss}

We will now turn to our second main result, namely that---under rather general assumptions which we will detail below---Gaussian 
states defined on slabs of a finite width cannot be used as perfect 
primitives for resources for measurement-based quantum computing, even if non-Gaussian measurements are allowed for:
Any influence of local measurements will again exponentially decay with the distance. 

More specifically, we will first show that a one-dimensional GPEPS cannot constitute 
a quantum wire in the sense of the definition of Subsection \ref{sus:MBQCgauss} extended to arbitrary measurements. 
This already covers all kinds of sophisticated encodings that can be carried by a single quantum wire, including 
ideas of ``encoding qubits in oscillators'' \cite{encoding}. 
We will then discuss the situation when an entire cubic slab of constant width is being used to encode a single 
quantum logical degree of freedom, and still find that the fidelity of transport 
will still decay exponentially. Not even using many modes and coupled quantum wires, 
possibly employing ideas of distillation, this obstacle can be overcome with local measurements alone.
That is to say, we show that Gaussian states can not be uplifted to serve as perfect universal resource
states by measurements on finite slabs alone: Frankly, the finite squeezing present in the initial resources---although 
the state being pure and known---has 
to be treated as a faulty state and some full machinery of {\it fault tolerance} \cite{faulttolerantnielsen,faulttoleranttraussendorf}, which 
yet has to be developed for this kind of system, necessarily has to be applied 
even in the absence of errors.  This quite severely contrasts with other limitations known for Gaussian quantum states. 
For example, while the distillation of entanglement 
is not possible using Gaussian operations alone, non-Gaussian operations help to overcome this 
task \cite{opticaldistillation}. 

%%%%%%%%%%%%%%%%%%%%%%%

\begin{figure}
\begin{center}
\includegraphics[width=6.6cm]{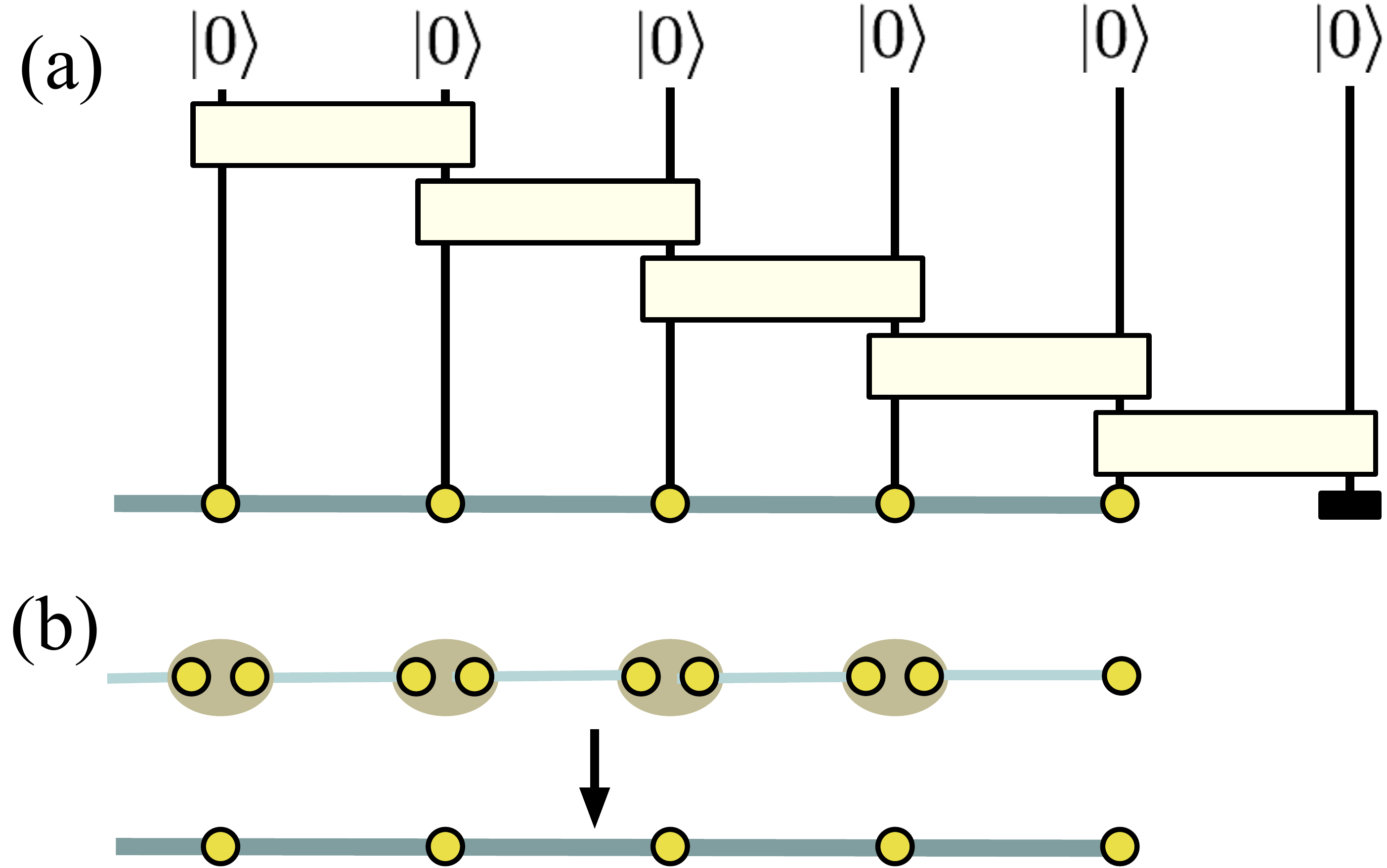}
\caption{\label{fig:sequential}(a) Sequential preparation of a Gaussian MPS state: Each line represents a mode of a unitary tensor 
network, whereas each box stands for a Gaussian unitary. For a suitable choice of the Gaussian unitaries, the resulting state is 
a Gaussian cluster state as being prepared in the valence bond construction (b).}
\end{center}
\end{figure}

\subsection{Sequential preparation of one-dimensional Gaussian quantum wires} 
\label{sus:sequential}

In order to state the statement, we first have to introduce another equivalent way of defining Gaussian PEPS or specifically
Gaussian MPS in one dimension: It is easy to see that a Gaussian MPS with state vector $|\psi\rangle$ of $N$ modes
can be prepared as 
\begin{equation}\label{seqe}
	|\psi\rangle =\langle \omega|_{N+1}\prod_{j=1}^N U^{(j,j+1)}|0\rangle^{\otimes (N+1)},
\end{equation}
with identical Gaussian unitaries $U^{(j,j+1)}$ supported on modes $j,j+1$, depicted as grey boxes in Fig.\ \ref{fig:sequential}.
This follows immediately from the original construction of Ref.\ \cite{MPStheoryhard}, see also Ref.\ \cite{MPStheoryeasy},
translated into the Gaussian setting. A detailed study of sequentially preparable infinite-dimensional quantum systems
with an infinite or finite bond dimension will be presented elsewhere.

\subsection{Impossibility of transport by non-Gaussian measurements in one 
dimension: General considerations}\label{general}

We will start by stating the main observation here: Frankly, even under general non-Gaussian measurements, 
transport along a 1D chain is not possible. We will refer both to the notions of localizable entanglement and the 
 {\it probability of transport}: This is the average maximum probability to 
 recover an unknown input state in a fixed subspace $S$ 
 of dimension of at least $\dim(S)\geq 2$ which has been transported through the wire: 
 Specifically, one asks for the maximum average success probability of a POVM applied to the output
 of the wire that leads to the
identity channel up to a constant, where the average is taken with respect to 
all possible outcomes when performing local measurements when transporting along the wire. 
We will see that  this probability will decay exponentially with the distance between the input and the output site.

This decay follows regardless of the encoding chosen. Note that by no means we require logical information 
to be contained in a certain fixed logical subspace along the computation: Only in the first and last steps---when 
initially encoding quantum
information or coupling to another logical qubit---we ask for a fixed subspace. This logical subspace is allowed
to even 
stochastically fluctuate along the computation dependent on 
measurement outcomes that are obtained in earlier steps of the computation. 

\begin{observation}[Impossibility of using Gaussian 1D chains as quantum wires under general measurements]
\label{the:onedgeneral}
Let $G$ be a one-dimensional GPEPS. Let $S$ be either $S={\cal H}$ or a subspace thereof.
Then the probability of transport between 
any two sites $A$ and $B$ of the wire satisfies
\begin{equation}
	p\le c_3 e^{-d(A,B)/\xi_3}
\end{equation}
for suitable constants
 $c_3,\xi_3>0$. This implies that for any subsets of sites $E_A$ and $E_B$
 and for fixed local subspaces, the entanglement between 
$E_A$ and $E_B$ that can be achieved by arbitrary local
measurements on all sites except those contained in $E_A$ and $E_B$ is necessarily
exponentially decaying in $d(E_A,E_B)$. This also means that for any two sites 
$A$ and $B$,
\begin{equation}
	E_S(A,B)\le c_4 e^{-d(A,B)/\xi_3}
\end{equation}
for some $c_4>0$ are constants, even if arbitrary local measurements are taken into account. 
\end{observation}

We now proceed in two steps. First, it is shown that there exists no subspace $S\in {\cal H}$ of dimension at least $\dim(S)\geq 2$ such that $V_j$ can be chosen to be unitary, for all $j$ for which $p_j>0$ and
\begin{equation}\label{SeqProp}
	\langle \eta_j |U |\psi\rangle|0\rangle = p_j^{1/2} V_j|\psi\rangle
\end{equation}
for all $|\psi\rangle\in S$ where all$U$ is the Gaussian unitary of the sequential preparation in Eq.\ (\ref{seqe}), 
where the index of the mode, and also any label of 
tensor factors, is suppressed, see Fig. \ref{fig:tensors}. $\{|\eta_j\rangle\}$ is an
orthonormal basis of ${\cal H}$, $j$ labeling the respective outcome of the local measurement,
possibly a continuous function. 
Because the computational subspace $S$ is allowed to vary during the 
processing but must be invariant for the computation as a whole, we have to consider all $N$ 
steps of the sequential preparation and all measurements together. For reasons of 
simplicity, yet, we first present the argument for a wire consisting of just two sites and extend it afterwards.  
We define the operator
\begin{equation}
\label{eq:M}
	M= U^\dagger (\id\otimes |0\rangle\langle 0|)U.
\end{equation}
and formulate the subsequent Lemma:

\begin{figure}
\begin{center}
\includegraphics[width=7cm]{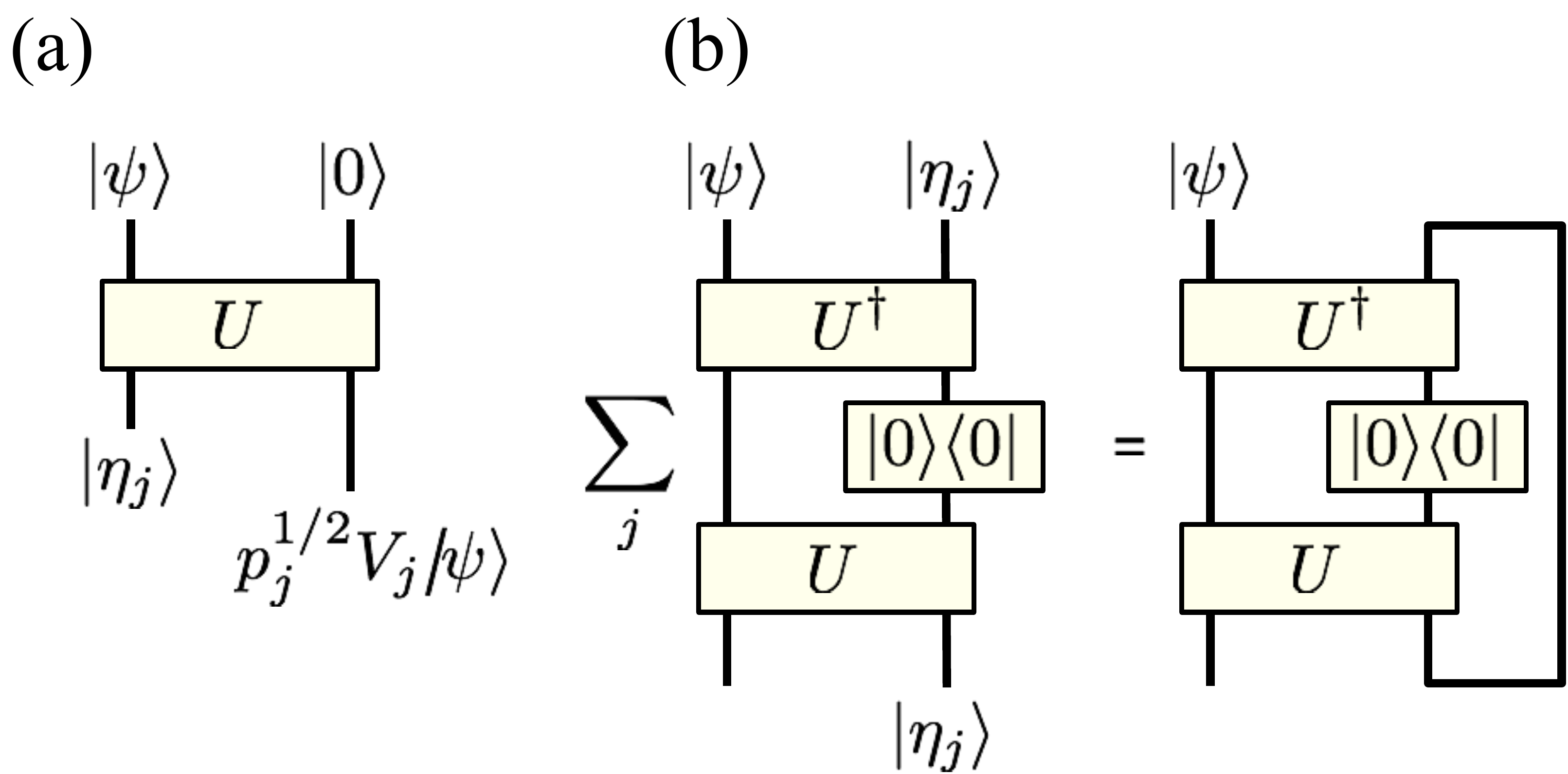}
\caption{\label{fig:tensors}(a) The network representing a single step of a sequential preparation of a Gaussian MPS, 
and (b) the tensor network representation of $\langle\psi|\langle0| U^\dagger(\id\otimes |0\rangle\langle0|)U|0\rangle|\psi\rangle$.
}
\end{center}
\end{figure}

\begin{lemma}[Conditions for non-decaying transport fidelity] 
\label{lem:thecond}
Necessary condition for Eq.\ (\ref{SeqProp}) to be satisfied is that 
\begin{equation}\label{thecond}
	 \langle \psi|\langle\eta_j|M|\psi\rangle|\eta_j\rangle= p_j
\end{equation}
for all $j$ and all $|\psi\rangle \in S$, with $\sum_j p_j=1$ and $\{|\eta_j\rangle\}$ forming a complete 
orthonormal basis of ${\cal H}$.
\end{lemma}
To see this note that the fact that Eq.\ (\ref{SeqProp}) holds true for each $j$ for any $|\psi\rangle \in S$ means 
that 
\begin{equation}
	P_{S}\langle \eta_j |U |0\rangle P_S = p_j^{1/2} P_S,
\end{equation}
where $P_S$ denotes the projection onto $S$. Using completeness of $\{|\eta_j\rangle\}$,
\begin{equation}
	\sum_j |\eta_j\rangle\langle\eta_j| = \id.
\end{equation}
a moment of thought reveals that for any $|\phi\rangle \in S^\perp$, the latter denoting the orthogonal complement of
$S$, one has that
\begin{equation}
	P_S\langle \eta_j |U |\phi\rangle|0\rangle = 0.
\end{equation}
What is more,
\begin{equation}
	\label{eq:eq0}
	\langle\phi |\langle \eta_j |U|0\rangle P_S= 0,
\end{equation}
again for all $|\phi\rangle \in S^\perp$.
This yet means that, see Fig.\ \ref{fig:tensors},
\begin{equation}
	\label{eq:eq1}
	 \langle \psi|\langle\eta_j|U^\dagger (\id\otimes |0\rangle\langle 0|)U|\psi\rangle|\eta_j\rangle = 
	  \langle \psi|\langle\eta_j|M|\psi\rangle|\eta_j\rangle= p_j,
\end{equation}
which proves Lemma \ref{lem:thecond}. Summing now over all measurement outcomes $j$ in 
Eq.\ (\ref{eq:eq1}) which is the same as performing the 
partial trace, see Fig.\ \ref{fig:tensors}, with respect to the second mode, we obtain
\begin{equation}
	 \langle \psi| \text{tr}_2\left(U^\dagger (\id\otimes |0\rangle\langle 0|)U\right)|\psi\rangle = 1,
\end{equation}
which in turn implies, together with the above that
\begin{equation}
	\label{eq:eq2}
	P_S\text{tr}_2\left(U^\dagger (\id\otimes |0\rangle\langle 0|)U\right)P_S=P_S.
\end{equation}
But this in turn means that the Gaussian operator $\text{tr}_2(U^\dagger (\id\otimes |0\rangle\langle 0|)U)$
has at least two spectral values that are identical. Now the only possibility for a Gaussian operator 
to have two equal, non-zero spectral values is to 
have a flat spectrum which corresponds to 
an operator which is not of trace-class (related to ``infinite
squeezing'' and ``infinite energy'' which was excluded due to the restriction to proper quantum 
states with finite energy).

We now extend the argument to a wire of arbitrary length. For this aim we denote the measurement basis on the 
$k$-th site by $\{|\eta^{(k)}_j\rangle\}$ and the corresponding probabilities by $p_j^{(k)}$. The definition (\ref{eq:M}) is generalized to
\begin{equation}
	{M}=\left(U^\dag(\id\otimes|0\rangle)\right)^N\left((\langle 0|\otimes\id)U\right)^N .
\end{equation}
Condition (\ref{SeqProp}) becomes
\begin{equation}\label{SeqProp2}
	\left(\otimes_k\langle \eta_j^{(k)}|\right)U^{\otimes N}|\psi\rangle|0\rangle^{\otimes N} =
	\prod_k (p_j^{(k)})^{1/2}V_{j}^{(k)}|\psi\rangle,
\end{equation}
where $\prod_k V_{j}^{(k)}$ is unitary for all sequences 
of measurement outcomes and, furthermore, acts trivially on $S^\perp$. 
Modifying also Eqs.\ (\ref{eq:eq0}), (\ref{eq:eq1}), and (\ref{eq:eq2}) in a similar manner 
and using the completeness of the $N$ measurement bases 
$\{|\eta^{(k)}_j\rangle\}$, we find that for Eq.\ (\ref{SeqProp2}) to hold, the Gaussian operator 
$O={\rm tr}^N (M)$, where ${\rm tr}^N$ denote the $N$-fold partial trace (or suitable tensor contraction), 
has two equal spectral values which is not possible as mentioned above and, thus, the first step of the 
proof is complete. 

\subsection{Impossibility of transport by non-Gaussian measurements in one 
dimension: Proving a gap}\label{gap}

In a second step we show now that Observation \ref{the:onedgeneral} holds if Eq.~(\ref{SeqProp}) is not fulfilled. 
The problem of recovering an unknown state after propagation through the wire is equivalent to the one of 
undoing a non-unitary operation. Obviously,
it is a fundamental feature of quantum mechanics that it is not possible to implement a 
non-unitary linear transformation in a deterministic fashion. Since one does not 
have to correct for a non-unitary operation in each step, however,
the technicality of the argument is related to the fact that we only
have to undo an entire word of non-unitary Kraus operators once.

Assume that we aim to use our wire for the transport of a single pure qubit. 
After $N$ steps of transport it will still be pure, 
but in general distorted due to the application of some non-unitary operator 
\begin{equation}
	V_J=V^{(N)}_{j_N}\dots V^{(1)}_{j_1}
\end{equation}	
where $J=(j_1,\dots, j_N)$ is an index reflecting the entire 
sequence of measurement outcomes on the $N$ lattice sites. 
To recover the initial state, one has to apply a 
$X_J$ such that 
\begin{equation}
	X_JV_J=c_J\id
\end{equation}	
 with $c_j\in \cc$.  The success probability of this recovery-operation, averaged over all measurement outcomes, 
 is nothing but the {\it probability of transport}. It will 
 decay exponentially in $N$ whenever for any $k$ at least a single 
 $V^{(k)}_{j_k}$ is not unitary. The maximal average 
 probability to undo random words $V_J$ of Kraus operators is found to be
\begin{eqnarray}
	p_N= \text{max} \, & \text{tr}(X_J V_J \rho V_J^\dagger X_J^\dagger),\label{eq:maximalpJ}
\\
	\text{subject to} \, & X_J^\dagger X_J= \id,\\
		&\, X_J V_J = c_J \id.
\end{eqnarray}
A moment of thought reveals that this probability of transport is then found to be
%Averaging (\ref{eq:maximalpJ}) over all measurement outcomes yields (note the cancellation of the dependency on the input state)
\begin{equation}\label{eq:FM}
	p_N=  \sum_J \lambda_1((V_J^\dagger V_J)^{-1})^{-1} =  \sum_J\lambda_n (V_J^\dagger V_J),
\end{equation}
where $\lambda_1$ ($\lambda_n$) denotes the largest (smallest) eigenvalue.

To show that Observation \ref{the:onedgeneral} is true if $V_k$ is not proportional to a unitary 
matrix for at least one $k$ can be shown by induction. Denoting, again, the operators applied 
by the measurements on the first $N$ sites by $V_J$ 
and the corresponding operators for site $N+1$ by $\{W_j\}$, we get from Eq.\ (\ref{eq:FM}) 
\begin{equation}
\label{eq:pNp1}
p_{N+1}=\sum_{J,j}\lambda_n(V_J^\dag W_j^\dag W_j V_J).
\end{equation}
Before we proceed, we note that it is possible to assume that all $W_j$ and $V_j$ are effective $2\times 2$ matrices, 
corresponding to the situation where the computational subspace $S$ does not change. 
If this is not the case, one can account for the fluctuation of the computational 
subspace by replacing $V_j\mapsto U_jV_j$ (and performing an analogous 
replacement for $W_j$) with a suitable 
unitary $U_j$. All arguments that follow will not depend on the choice of 
this unitary $U_j$.
Key to the exponential decay is a Lemma that will be 
proven in Appendix \ref{app:matrixanalysis}.

\begin{lemma}[Bound to eigenvalues of the sum of $2\times 2$ matrices]
\label{lem:matrixanalysis}
For any positive $A,B\in\cc^{2\times2}$ with $[A,B]\neq 0$ there exists a $\delta>0$ such that 
\begin{equation}
\label{eq:matrixanalysis}
\lambda_2(A+B)\ge\lambda_2(A)+\lambda_2(B)+\delta\,.
\end{equation}
\end{lemma}

If there exists at least one pair $(i,j)$ for which
\begin{equation}
	[W_i^\dag W_i,W_j^\dag W_j]\neq 0,
\end{equation}	
then also
\begin{equation}
	[V_J^\dag W_i^\dag W_i V_J, V_J^\dag W_j^\dag W_j V_J]\neq 0,
\end{equation}
and
we can apply Lemma \ref{lem:matrixanalysis} directly to Eqs.\ 
(\ref{eq:pNp1}). If in contrast
\begin{equation}
	[W_i^\dag W_i,W_j^\dag W_j]=0 
\end{equation}	
for all pairs $(i,j)$, all $W_i^\dag W_i$ can be  simultaneously diagonalized. 
This means that we 
can---without loss of generality---assume that 
\begin{equation}
	W_i^\dag W_i=\text{diag}(\xi_{i},\zeta_{i}) .
\end{equation}	
Because a 
non-unitary $W_i$ exists by assumption, $\min \{| \xi_{i}  - \zeta_{i}|:i=1,2\}>0$. 
In both cases we are provided with a $\nu<1$ 
such that
\begin{equation}
\label{eq:pNp1N} 
p_{N+1}\leq  \nu\sum_J\lambda_2(V_J^\dag V_J)= \nu p_N,
\end{equation}
where we have used the completeness relation 
\begin{equation}
	\sum_j W_j^\dag W_j=\id. 
\end{equation}
This observation gives rise to the anticipated gap that 
proves the exponential decay of the probability of transport and, therefore, to 
Observation \ref{the:onedgeneral}. The exponential decay of the subspace localizable entanglement follows directly: If there 
was a non-decaying localizable entanglement, this could be used to transport with high 
recovery probability in contrast to what we have shown. 
If this was not the case, one could use the wire to distribute entanglement which is obviously not possible.

\subsection{Impossibility of transport by non-Gaussian measurements in one 
dimension: Concluding remarks}\label{concluding}

Note, finally, that even though we have presented Observation \ref{the:onedgeneral} for local projective measurements---which
suits the paradigm of measurement-based computing---the argument obviously holds true for 
POVM measurements. The proof is completely analogous, with $\sum_j |\eta_j\rangle\langle\eta_j|=\id$ being
replaced by a more general resolution of the identity.

This argument shows that one-dimensional GPEPS cannot be used as quantum wires even when allowing for arbitrary 
non-Gaussian local measurements. Note that for this argument to hold, completeness of the measurement 
bases are indeed necessary: For single outcomes, the condition of the output being up to a constant unitarily 
equivalent to the input can well be achieved also for matrices having a different structure; but then, one 
cannot make sure that this is true for each outcome $j$ of the measurement. This, however, is required in order to faithfully transport quantum information. 
If we allow for a finite rate of failure outcomes $j$ in individual steps, then the overall probability of success will 
asymptotically again become zero at an exponential rate.

\subsection{Gaussian cluster states under arbitrary encodings and in higher-dimensional lattices}
\label{sus:locnogauss}
One might wonder whether this limitation can be overcome if a large number of physical modes of a higher-dimensional
lattice are allowed to carry logical information. 
The same argument, actually, can be applied to a
$k\times k\times \dots\times k\times n$ cubic slab, as a subset of a $D$-dimensional cubic lattice,
where one aims at transporting along the last dimension, with
local measurements at each site. In fact, contracting any dimension except from the last---so summing over all 
joint indices---one arrives at a Gaussian MPS, with a bond dimension that is exponential in $k$. This, yet, is a constant.
This situation is hence again covered by a Gaussian MPS, as long as one allows for more than one physical modes
and more than one virtual modes per site. Since the above argument in Subsection \ref{general} does not make
use of the fact that we only have a single virtual and physical mode per site, only that now $|0\rangle^{\otimes (k^{(D-1)})}$
are being fed into the sequential preparation. 

\begin{observation}[Exponential decay of subspace localizable entanglement in a higher-dimensional lattice]
\label{the:onedhigherdim}
Let $G$ be a one-dimensional GPEPS and $A$ and $B$ two sites in a $k\times k\times \dots\times k\times n$ slab as a
subset of a 
$D$-dimensional cubic lattice, and denote with $i,j$ the last coordinate of sites $A$ and $B$. Then
\begin{equation}
	E_S(A,B)\le c_4 e^{-d(i,j)/\xi_4}
\end{equation}
where 
$c_4,\xi_4>0$ are constants, even if arbitrary local measurements are taken into account. 
\end{observation}

\begin{figure}
\begin{center}
\includegraphics[width=5.2cm]{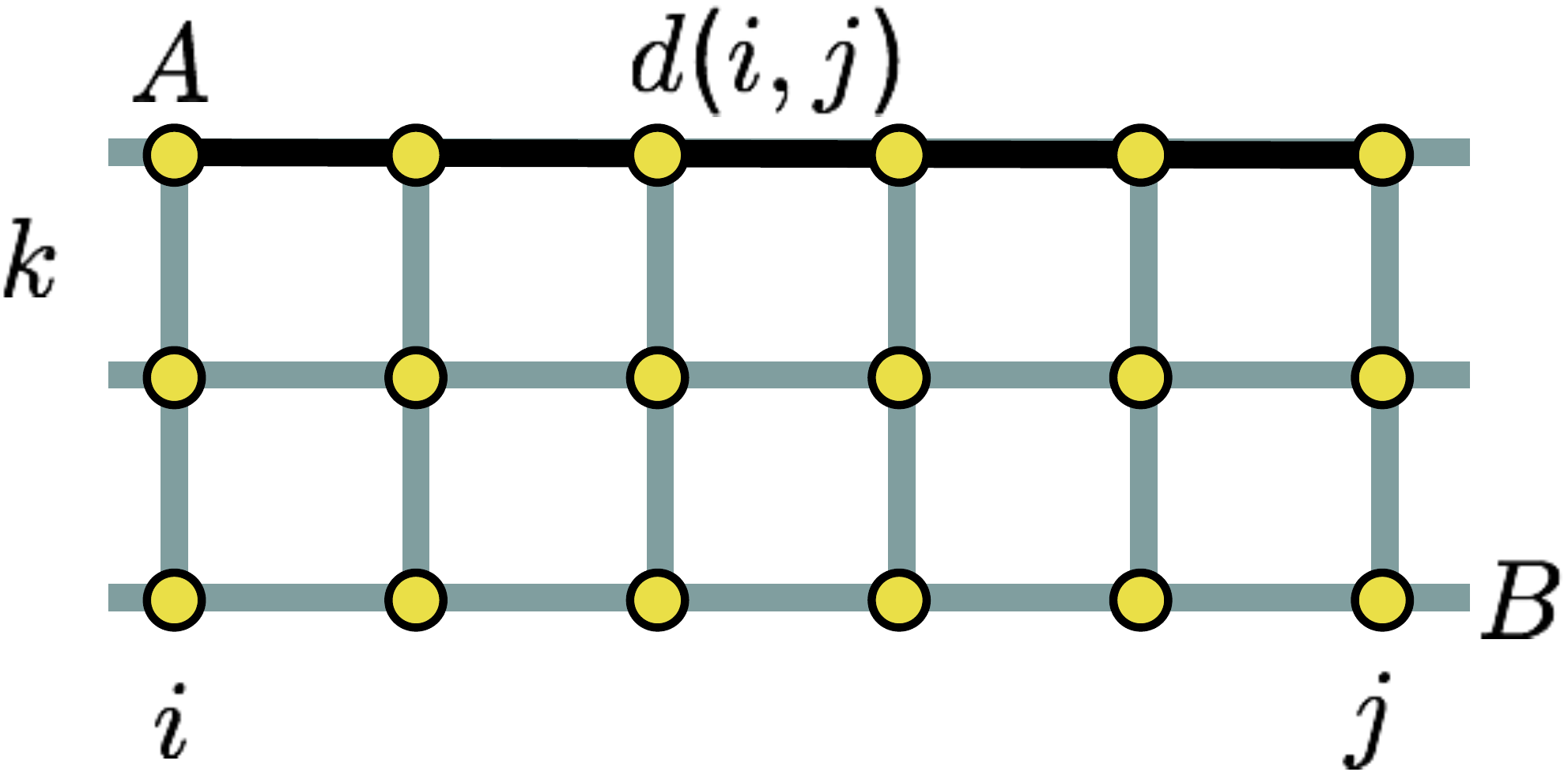}
\caption{\label{fig:slap}A slab of a $k\times n$-lattice, aiming at using the second dimension as a quantum wire
for quantum computation. Again the probability of transport between $A$ and $B$ decays exponentially
with the distance along the last dimension.}
\end{center}
\end{figure}

So even encodings in larger-dimensional Gaussian cluster states do not alter the situation that one cannot
transport along a given dimension, if one wants to think of such slabs as perfect primitives being
used in a universal quantum computing scheme.

\subsection{Role of error correction and fault tolerance}

The above Observations \ref{the:generalgraph} and \ref{the:onedhigherdim} 
show that under mild conditions, Gaussian cluster states {\it not be used as} or 
{\it made almost perfect resources} by local measurements alone. This constitutes a 
significant challenge for measurement-based quantum computing with Gaussian cluster states,
but does not rule our this possibility. In this subsection, we briefly
comment on ways that possibly allow to overcome the limitations identified here. 

Clearly, it is very much conceivable that this observation may again be overcome by concatenated encoding in fault-tolerant
schemes, effectively in slabs the width of which scales with the length of the computation: Rather at the level of 
finite encodings, the resource cannot be uplifted to a perfect resource. 
The situation encountered here---having pure Gaussian states---has hence some 
similarity with {\it noisy finite-dimensional cluster states} built with {\it imperfect operations} \cite{faulttolerantnielsen,faulttoleranttraussendorf}. 
Considering the preparation of the quantum wire and the transport by local measurements 
as a sequence of teleportations with not fully entangled resources, this means that every step adds a 
given amount of noise to the quantum information. In finite-dimensional schemes, if this noise corresponds to an error rate below the 
fault-tolerant a nested encoding with an error-correction code allows to perform computations. The size of the code grows polynomially with the size of the circuit one wishes to implement. In addition to this {\it intrinsically} error, any physical implementation will, of course, 
also suffer from {\it experimental} errors which must be also compensated by error-correction schemes. 
Thus, the {\it combined error} rate must be below the fault-tolerant threshold. It is therefore
possible that once recognizing all finite squeezings as full quantum errors---which has to be done in the light of the
results of the present work---and using suitable concatenated encodings over polynomially many slabs, 
that there exists a finite squeezing allowing for full universal quantum computation with eventual polynomial overhead.
The question whether such schemes can be composed, or ones where suitable polynomially sized
complex structures are ``pinched'' out of a large lattice---that are universal remains a
challenging interesting open question. 

\subsection{Ideas of percolation}

One possible way forward in this direction to achieving a fully universal resource under 
local non-Gaussian measurements would be to think of first performing local measurements at each site, aiming at
filtering an imperfect qubit, $\cc^2$-cluster from a Gaussian cluster state. Ideally, one would arrive at
the situation on, say, a cubic lattice of some dimension, where one can extract a {\it graph state} \cite{Hein}
corresponding to having an edge between nearest neighbors with some finite probability.
If this probability $p_s$ is sufficiently high---larger than the appropriate threshold
for {\it edge percolation}---and if one can ensure suitable independence, 
an asymptotically perfect cluster on a renormalized lattice can be 
obtained \cite{percolation,percolationlong,percolation2}. When trying to identify such percolation schemes, one 
does not have to rely on classical percolation schemes, but can also make use of more general repeater-type 
schemes of Ref.\ \cite{Acin}, then referred to as {\it quantum percolation} (see also Ref.\ \cite{percolationlong}). 
To identify such maps, either classical or quantum, yet, appears to be a very challenging task.

One might also ask whether the TMSS bonds as such can be transformed into suitable maximally
entangled pairs of $\cc^2\otimes \cc^2$ systems. This, however, clearly is the case.
Again applying a result for finite-dimensional systems to infinite dimensional ones by making
use of appropriate nets of Hilbert spaces, one finds that given a state
vector $|\psi_\lambda\rangle$ of a TMSS of some squeezing parameter $\lambda>0$, the transformation
$|\psi_\lambda\rangle$ to $(|0,0\rangle + |1,1\rangle)/\sqrt{2}$ is possible with a generalized local filtering
on $A$ only, together with a suitable unitary in $B$, with a probability of success of \cite{nielsen,majo}
\begin{equation}
	p = \min(1, 2 (1-\lambda^2)).
\end{equation}
Hence, whenever $\lambda\ge1/\sqrt{2}$, this transformation can be done deterministically. This has interesting consequences for 
quantum repeaters. The protocol performing the transformation
\begin{equation}
\label{eq:belltrafo}
	|\psi_\lambda\rangle_{A,B}\mapsto\frac{1}{\sqrt{2}}\left(|0,0\rangle+|1,1\rangle\right)
\end{equation}
can be implemented by combining $A$ with an ancillary system $C$, performing a joined unitary transform on $A,C$, measuring $C$ and applying another unitary gate on $B$ classically conditioned on the measurement result. 

But even if $\lambda<1/\sqrt{2}$ one can still distill a resource from a collection of TMSS distributed on a graph, 
performing an argument involving percolation here. 
This yet merely shows that Gaussian states as such can be 
resources for information processing. Most importantly, this is not the resource anticipated, so not the actual
GPEPS, but a collection of suitable TMSS. 
Then, non-Gaussian PEPS projections
cannot be implemented with linear optics without an massive overhead. 
Finally, an eventually created qubit cluster state would be obtained in \textit{single-rail} representation where measurements 
in the superposition bases, which are needed for the actual computation, are experimentally very difficult and 
require additional photons. So the question of actual universality of the
Gaussian cluster state under all fair meaningful ways of defining a set of rules
remains an interesting challenging question.

%%%%%%%%%%%%%%%%%%%%%%%%%%%%%%%%%%%%%%%
\subsection{Remarks on one-dimensional Gaussian quantum repeaters}
 
We finally briefly reconsider the question of a quantum repeater setting based on general non-Gaussian operations.
Above, we have shown that it is not possible to obtain a finitely entangled state for an arbitrary long one-dimensional GPEPS. Yet, what is also true at the same time is that a sequential 
repeater scheme based on sufficiently entangled 
TMSS {\it before} the PEPS projection does yield a non-decaying entangled bond between the end points. That is,
using only projective local measurements on each of the sites, one can transform a collection of distributed TMSS 
in a 1D setting into a maximally entangled qubit pair shared between the end sites. In order to show this, it suffices
to revisit the situation for three sites, as the general statement on $N$ sites follows immediately by iteration.

Now consider the quantum repeater setting and assume for simpli/city that we already have a qubit Bell-pair 
$|\phi\rangle_{A,B_1}= (|0,0\rangle+ |1,1\rangle)/\sqrt{2}$ 
which we want to swap through a TMSS $|\psi_\lambda\rangle_{B_2,C}$ with $\lambda\ge1/\sqrt{2}$. 
We can use the higher, unoccupied Fock-levels of 
the state vector $|\phi\rangle_{A,B_1}$ as an ancilla to transform $|\psi_\lambda\rangle_{B_2,C}$ according to Eq.\
(\ref{eq:belltrafo}). As the final unitary on $C$ after LOCC with one-way classical communication
does not change the entanglement, we can as well omit it. As the unitary, the ancilla-measurement and the final Bell-measurement on 
$B_1,B_2$ are equivalent to a single projective measurement on $B_1,B_2$, it is possible to  
swap entanglement through an physical TMSS perfectly. Needless to say, this will be a highly non-Gaussian
complicated operation, and will not overcome the limitation of Gaussian cluster states discussed above.

\section{Discussion and summary}

In this article, we have have assessed the requirements to possible architectures
when using Gaussian states as resources for measurement-based quantum computing and for entanglement distribution by 
means of quantum repeater networks. Using a framework of Gaussian PEPS, we have shown that 
under Gaussian measurements only, the localizable entanglement decays exponentially with the distance on arbitrary graphs. 
This rules out the possibility to process or even transport 
quantum information with Gaussian measurements only.

The above results also show that Gaussian cluster states---under mild conditions
on the encoding of logical information in slabs, rather than having general encodings in the entire lattice---can not be 
used as or made perfect universal resources for measurement-based quantum computation. 
No information can be transmitted beyond a certain influence region,
and hence, no arbitrarily long computation can be sustained. Now, if one allows for larger energy, and hence larger two-mode squeezing,
in the resource states, this influence region will become larger. In other words, small-scale implementations as proof-of-principle 
experimental realizations of such an idea will be entirely unaffected by this: Any state with finite energy will constitute some
approximation of the idealized improper state having infinite energy, and its outcomes in measurements will approximate the
idealized ones. Only that with this state, one could not go ahead with an arbitrarily long computation. This observation
shows that Gaussian cluster states are fine examples of states that eventually allow for the demonstration of the functioning
of a continuous-variable quantum computer, possibly realized using the many modes available in a {\it frequency comb}
\cite{gaussiancluster,gaussiancluster2,gaussiancluster3}. 

Also, we have discussed the requirements for fault tolerance and
quantum error correction for such schemes, yet to be established, in that any finite squeezings essentially have to be 
considered full errors in a concatenated encoding scheme.
This work motivates such further studies of fault-tolerance of systems with a finite-dimensional logical encoding
in infinite-dimensional systems. But it also strongly suggests
that it could be a fruitful enterprise to further at alternative CV schemes, not directly involving Gaussian states, but other relatively
feasible classes of states, such as coherent superpositions of a few Gaussian states like the so-called {\it cat states}, 
which have turned out to be very useful within another computation paradigm \cite{cats}.
We hope that this article can contribute to sharpen the needs that any architecture eventually
needs to meet based on the interesting idea of doing quantum computing by performing
local measurements on Gaussian or non-Gaussian states of light. 
  
 \subsection{Acknowledgments}

We acknowledge interesting discussions with 
M.\ Christandl,
S.~T.\ Flammia, 
D.\ Gross,
P.\ van Loock,
N.\ Menicucci, and 
T.\ C.\ Ralph.
This work has been supported by the EU 
(COMPAS, QAP, QESSENCE, MINOS) and the EURYI scheme.

\begin{appendix}
\section{Proof of Lemma \ref{lem:matrixanalysis}
\label{app:matrixanalysis}}
Let $A,B \in\cc^{2\times 2}$ with $A,B\ge 0$. We set
\begin{equation}
	c= \frac{ \| A^{1/2} B^{1/2}\|^2}{\|A\|\|B\|}.
\end{equation} 
The inequality $c\le 1$ follows directly from the submultiplicativity of he operator 
norm while equality holds if and only if $A$ and $B$ commute. 
Rewriting
\begin{eqnarray}
	\lambda_n(A + B) &=& \text{tr}(A + B)- \lambda_1(A + B)\nonumber\\
	&=& \text{tr}(A + B)- \| A + B\|,
\end{eqnarray}
we can now use a sharpened form of the triangle inequality for the operator norm of $2\times 2$-matrices in Ref. \cite{Triangle} to obtain
\begin{eqnarray}
	\lambda_2(A + B)
	&=& \text{tr}(A + B)- \| A + B\|\\
	&\geq & \text{tr}(A + B)-\frac{1}{2}(\|A\|+\|B\|)\nonumber\\
	&+& \frac{1}{2}
	\left(
	(\|A\|-\|B\|)^2 + 4 \| A^{1/2} B^{1/2}\|^2
	\right)^{1/2}\nonumber.
\end{eqnarray}
 If now $c<1$, then there exists a $\delta>0$ such that
\begin{eqnarray}
	\lambda_2(A + B)
	&\geq& \text{tr}(A + B)\\
	&-&\left(
	(\|A\|-\|B\|)^2 + 4 \| A\| \|B\|
	\right)^{1/2}+\delta\nonumber\\
	&=&  \text{tr}(A + B) - (\|A\|+ \|B\|)+\delta\nonumber\\
	&=&  \lambda_2(A) + \lambda_2(B)+\delta
\end{eqnarray}	
which proves Lemma \ref{lem:matrixanalysis}.

\end{appendix}

\end{document}